\documentclass[useAMS,usenatbib]{mn2e}
\usepackage{willman_macros}
\usepackage{psfig}

\begin{document}

\title[The Observed and Predicted Spatial Distribution of Milky Way Satellite Galaxies]{The Observed and Predicted Spatial Distribution of Milky Way Satellite Galaxies}

\author[B. Willman \etal]{B~ Willman$^{1,2}$ \thanks{Email: beth.willman@nyu.edu}, F.~Governato$^{2,3}$ \thanks{Brooks Fellow},  J.~J.~Dalcanton$^2$ \thanks{Alfred P. Sloan Research Fellow}, D.~Reed$^{2,4}$ and T.~Quinn$^2$ \\
$^1$Center for Cosmology and Particle Physics, Department of Physics, New York University, 4 Washington Place, New York, NY 10003, USA  \\
$^2$Department of Astronomy, University of Washington, Box 351580, Seattle, WA 98195, USA \\
$^3$Osservatorio Astronomico di Brera (INAF), Via Brera 28, 20121 Milano, Italy\\
$^4$ICC, Dept. of Physics, University of Durham, Rochester Building, Science Laboratories, South Road, Durham DH1 3LE, UK}

\date{}             \pagerange{\pageref{firstpage}--\pageref{lastpage}}
\pubyear{2003}

\maketitle


\begin{abstract}

We review evidence that the census of Milky Way satellites similar to
those known may be incomplete at low latitude due to obscuration and
in the outer halo due to a decreasing sensitivity to dwarf satellites
with distance. We evaluate the possible impact that incompleteness has
on comparisons with substructure models by estimating corrections to
the known number of dwarfs using empirical and theoretical models.
Under the assumption that the true distribution of Milky Way
satellites is uniform with latitude, we estimate a 33$\%$
incompleteness in the total number of dwarfs due to obscuration at low
latitude.  Similarly, if the radial distribution of Milky Way
satellites matches that of M31, or that of the oldest sub-halos or the
most massive sub-halos in a simulation, then we estimate a total
number of Milky Way dwarfs ranging from 1 -- 3 times the known
population.  Although the true level of incompleteness is quite
uncertain, the fact that our extrapolations yield average total
numbers of MW dwarfs that are realistically 1.5 -- 2 times the known
population, shows that incompleteness needs to be taken seriously when
comparing to models of dwarf galaxy formation. Interestingly, the
radial distribution of the oldest sub-halos in a $\Lambda$CDM
simulation of a Milky Way-like galaxy possess a close match to the
observed distribution of M31's satellites, which suggests that
reionization may be an important factor controlling the observability
of sub-halos.  We also assess the prospects for a new SDSS search for
Milky Way satellites to constrain the possible incompleteness in the
outer halo.

\end{abstract}




\section{Introduction}
\label{sec:intro}

The currently favored $\Lambda$ + cold dark matter ($\Lambda$CDM)
cosmological model successfully reproduces many of the observed
large-scale properties of the Universe, including the properties of
the Cosmic Microwave Background recently observed by WMAP
\citep{spergel03}, the number, size and clustering of galaxy clusters
(e.g. \citealt{eke96,zehavi02}), and the evolution of galaxy cluster
counts \citep{rosati02}.  However, several major discrepancies between
the predictions of $\Lambda$CDM and the observed properties of the
Universe on small scales have presented challenges to the paradigm.
One outstanding challenge is that Cold Dark Matter models predict over
an order of magnitude more low mass, dark matter halos around the
Milky Way than the number of observed satellite dwarf galaxies.  This
discrepancy was first pointed out by \citet{kauffmann93}, and was
later confirmed by high resolution N-body simulations
\citep{klypin99,moore99,font01}.


Within the $\Lambda$CDM framework, a plausible explanation for the
discrepancy between the number of predicted sub-halos and observed
satellites lies within baryonic physics.  The fraction of dark matter
halos with $v_c \lesssim$ 50 km sec$^{-1}$ that may host a luminous
galaxy can be significantly reduced by reionization, feedback, and/or
tidal effects
(\citealt{dekel86,efstathiou92,thoul96,quinn96,bkw00,bkw01,benson02,susa03,dijkstra04},
among others).  These baryonic processes are difficult to
model. Therefore, the exact extent to which each effects the
present day luminosity of dark matter sub-halos remains uncertain.
A comparison of the total number and the radial distribution
of predicted visible satellites with that of the observed Milky Way
satellites may provide a test of the feasibility of particular models
(e.g. \citealt{taylor03,kravtsov04}).

A number of the above implementations of baryonic physics have been
able to reproduce the observed Milky Way dwarf
population. Unfortunately, existing comparisons are rendered less
meaningful by the uncertain completeness of the Milky Way dwarf
satellite population.  Due to incompleteness, the observed satellites
may not reflect the properties of the underlying population.  Models
that provide a good match to the current observations may, therefore,
actually underpredict the underlying population.

Past searches for Milky Way companions, although very successful,
suffer from unavoidable observational biases that could lead to an
undercounting of Milky Way satellites both at low Galactic latitudes
and at large ($>$ 100 kpc) distances (see \S3 for
discussion). Furthermore, the extent of these possible biases is not
well understood due to a lack of systematic analyses (however, see
\citealt{kleyna97} for a systematic analysis of their survey's
sensitivity).  \citet{willman02} are currently implementing a new search for
resolved Milky Way dwarf satellites in the Sloan Digital Sky Survey
data.  In contrast to past surveys, this search is sensitive to dwarfs
similar to and much fainter than any among the known population, at
any distance out to the Milky Way's virial radius.  The SDSS may thus
provide the means to evaluate the possibility of undercounting in the
outer halo.

In light of the new Willman et al survey, we reevaluate the evidence
for incompleteness to Milky Way satellites similar to those in the
known population and highlight the possible impact of such an
incompleteness on comparisons with substructure models. In \S2,
we describe the N-body cosmological simulation of a Milky Way like
galaxy that we use to compare with observations. In \S3, we review
observational evidence for bias in the census of Milky Way companions
and estimate the possible number of undetected galaxies similar to
those known, based on primarily observational arguments.  In \S4, we
compare the radial distribution of Milky Way dwarfs with that of both
the oldest and highest v$_c$ dark matter sub-halos of the simulated
galaxy.  We use this comparison both to demonstrate how well radial
distributions may be used to distinguish between models and to
underscore the possible impact of observational bias on such a
comparison.  Finally, in \S5 we estimate the number of dwarfs similar to the
known population that the Willman et al. (2002) survey could detect
and still be consistent with either of the two models that we
consider.

\section{The Simulation}

\citet{reed03} recently simulated the formation of a Milky Way-like
disk galaxy in a $\Lambda$CDM Universe.  They performed a dark matter
(DM) only simulation of a Milky Way sized galaxy halo, using PKDGRAV
\citep{stadel01}. In the following, we use ``DMgal'' to refer to this
galaxy. The details of this simulation are in \citet{governato03}, but
we summarize its properties here.  They adopted $\Omega_0 = 0.3$,
$\Lambda = 0.7$, $h$ = 0.7, $\sigma_8 = 1$, and $\Gamma$ = 0.21, where
$\Gamma$ is the shape parameter of the power spectrum.
Table~\ref{tab:simulations} lists the main parameters of the
simulation at z = 0.  We use $\delta \rho$/$\rho$ $\sim$ 100 to define
the virial radius of the galaxy \citep{eke96}.

\begin{table}
\begin{tabular}{cccccc}
\hline
 & $R_{vir}$ & $M_{vir}$ & N$_{dark}$  & M$_{dark}$ & $\epsilon$ \\
 & kpc & M$_{\odot}$ & within $R_{vir}$ & M$_{\odot}$  & kpc \\ \hline
DMgal & 365 & 2.9 $\cdot$ 10$^{12}$ & 864,744 & 3.3 $\cdot 10^6$ & 0.5 \\ \hline
cl1c6 & 1700 & 2.9 $\cdot$ 10$^{14}$ & 4,568,456 &  6.3 $\cdot 10^7$ & 1.25 \\ \hline
\label{tab:simulations}
\end{tabular}
\caption{Simulation Data}
\end{table}

\begin{figure}
\centerline{\psfig{file=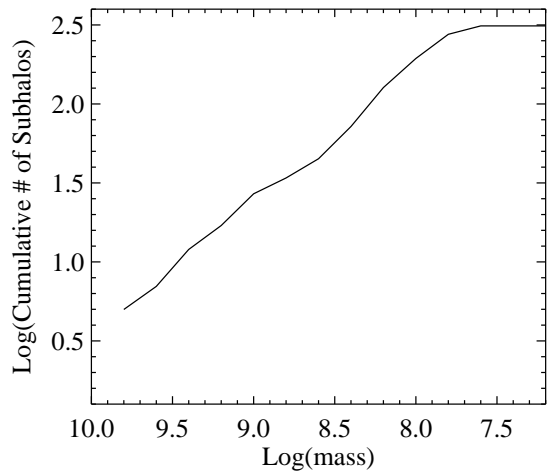,width=\hsize}}
\caption{The cumulative number of sub-halos within R$_{vir}$ as a function of mass in a dark matter only  $\Lambda$CDM simulation of a 3 $\cdot$ 10$^{12} M_{\odot}$ galaxy described in \citet{governato03}.}
\label{fig:numvsmass}
\end{figure}

To identify DMgal's sub-halos, we use the
SKID\footnote{http://www-hpcc.astro.washington.edu/tools/skid.html}
halo finding algorithm with a linking length of both 3 kpc and of 2
kpc.  We used the smaller linking length to include small halos that
are missed by the longer linking length.  Figure~\ref{fig:numvsmass}
shows the resulting cumulative number of dark matter sub-halos as a
function of mass.  The cumulative number of DMgal's sub-halos scales
roughly as $M^{-2}$ and does not flatten until masses below 10$^8$
M$_{\odot}$.

\begin{figure}
\centerline{\psfig{file=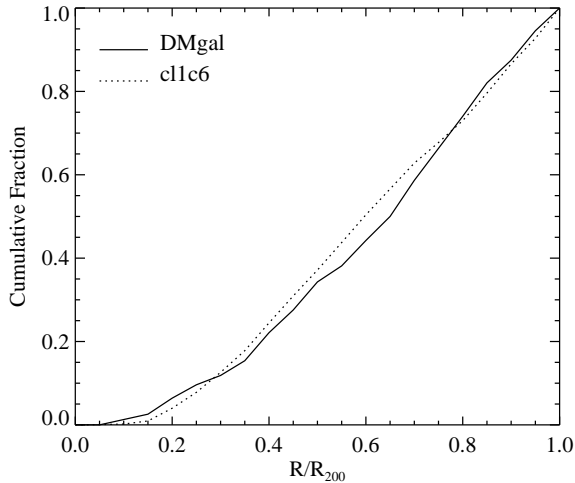,width=\hsize}}
\caption{The radial distribution of dark matter sub-halos, in the high
resolution, dark matter only simulation described in
\citet{governato03}, as compared to that of a higher resolution
simulation of a galaxy cluster \citep{reed03}. }
\label{fig:numvsdistmass}
\end{figure}

We compare the radial distribution of DMgal's sub-halos with that of a
higher resolution galaxy cluster simulation, `cl1c6', to ensure the
number of sub-halos at small radii is not resolution limited.  Table 2
includes the properties of cl1c6 from \citet{reed03}.
Figure~\ref{fig:numvsdistmass} shows that the two radial distributions
are consistent with each other, and also closely match that of the higher
resolution dark matter galaxy in \citet{stoehr03}, demonstrating
that the radial distribution of DMgal's sub-halos is not significantly
affected by overmerging.

\section{Observational Evidence For Incompleteness in the Milky Way Dwarf Galaxy Census}

There is observational evidence that the census of Milky Way
companions may be incomplete at low Galactic latitudes and at
Galactocentric distance $\gtrsim$ 100 kpc due to observational biases.
In this section, we discuss these possible incompletenesses and
crudely estimate a reasonable correction to the currently known number
of dwarfs.

\subsection{Incompleteness at Low Galactic Latitude}
\begin{figure}
\centerline{\psfig{file=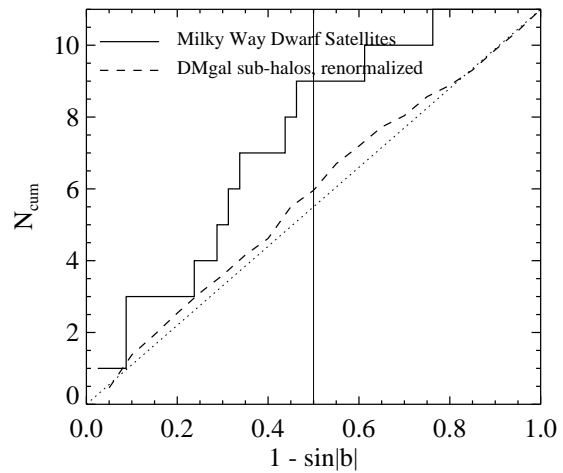,width=\hsize}}
\caption{The cumulative number distribution of Milky Way satellites as
a function of Galactic latitude, where low values of 1 - sin$|b|$ are
toward the Galactic poles.  The dotted line shows a uniform spatial
distribution of 11 dwarfs.  The dashed line shows the latitude
distribution of DMgal sub-halos, assuming that galaxy disks are
perpendicular to the major axis of their dark matter halos and
renormalized to the total number of Milky Way satellites.  The
vertical line shows where 50$\%$ of the cumulative distribution would
lie, for a uniform population. The asymmetry in the distribution
implies that 7 $\pm$ 2 satellites, similar to the known Milky Way
satellites, may lie undetected at low $b$.  Based on Figure 2b from
\citet{mateo98}.}
\label{fig:mwnumb}
\end{figure}

The increased extinction and stellar foreground toward the Galactic
disk severely limit the detectability of dwarfs that may lie at low
latitude.  This bias could account for the observed asymmetric
distribution of Milky Way satellites with latitude, pointed out by \citet{mateo98}.
Figure~\ref{fig:mwnumb}, based on Figure 2b in \citet{mateo98}, shows
the cumulative number distribution of the 11 Milky Way dwarf
satellites as a function of Galactic latitude.  If the true
distribution of dwarfs around the Milky Way is uniform with latitude,
then their cumulative number will increase linearly from the Galactic
poles with increasing 1 - sin$|b|$ \citep{mateo98}.  The dotted line
in Figure 2 thus shows the predicted distribution of a uniform
population of 11 dwarf satellites.  For reference, the solid line
shows where 50$\%$ of such a distribution would lie.

Assuming a uniform latitude distribution, the fact that 9 of the 11
known dwarfs have been detected at Galactic latitudes above the
expected 50$\%$ point implies a total of 18 $\pm$ 4 galaxies with
similar properties to the known dwarfs.  This number represents a
crude approximation of the effect of observational bias at low
latitude, as we ignore the possibility that Milky Way dwarfs are not
randomly distributed, but rather are distributed in 'dynamical
families' \citep{majewski94,fusi95,lyndenbell95,palma02}. 

Another uncertainty in the above estimate is that the intrinsic
distribution of dwarfs may not be uniform.  For example, \citet{k96}
found that M31's satellites follow an elongated distribution.
However, the 3 M31 satellites discovered since then decrease the
extent of the spatial asymmetry \citep{armandroff99}.  There is also
some observational evidence that the satellites of isolated disk
galaxies may be biased to lie at $|b|$ $>$ 30
(\citealt{holmberg74,zaritsky97,zaritsky99}).  However, it is unclear
how this ``Holmberg effect'' observed in isolated galaxies may
translate to galaxies in richer environments, such as the Local Group.
Furthermore, this effect has only been observed in satellites with d
$<$ 50 kpc \citep{holmberg74} or 300 $<$ d $<$ 500 \citep{zaritsky97},
and has only been reproduced in one published numerical simulation of
a disk galaxy \citep{penarrubia02}. \citet{knebe03} recently showed
that the orbits of simulated galaxy cluster sub-halos are biased to
lie along the major axis of the cluster, due to infall along
filaments.  They hypothesized that if galaxy disks are perpendicular
to the major axis of their dark matter halos, that the bias they
observe in simulated clusters may explain the Holmberg effect.

We computed the latitude distribution of DMgal's sub-halos to
investigate whether the non-uniformity seen in the
\citet{knebe03} cluster simulations is also seen in our galaxy
simulation.  We determined the latitude of each sub-halo assuming that
galaxy disks are perpendicular to the major axis of their associated
dark matter halos. To determine the shape and orientation of DMgal, we
used the {\it moments} command in
TIPSY\footnote{http://www-hpcc.astro.washington.edu/tools/tipsy/tipsy.html},
based on the technique described in \citet{katz91}.  The latitude
distribution of DMgal's sub-halos, overplotted on
Figure~\ref{fig:mwnumb}, does not have the asymmetry seen in that of
Milky Way satellites.  The fact that the latitude distribution of
DMgal's sub-halos is uniform shows that the \citet{knebe03} result is
not necessarily universal, and that Milky Way satellites are possibly
distributed uniformly with latitude.

The very recent discoveries of low latitude, low surface brightness
stellar structures around the Milky Way (Monoceros stream:
\citealp{newberg02,yanny03,rochapinto03,ibata03,crane03,martin04};
TriAnd: \citep{rochapinto04}) also lend strength to the interpretation that
the apparently asymmetric latitude distribution is at least partially
due to observational bias.  It is possible that each of these streams
may be associated with a low mass dwarf galaxy that is currently
undergoing tidal disruption.  However, a distinct core has not been
clearly detected in either system.  Such systems would not necessarily
have been identified as a 'dwarf galaxy' or 'dark matter halo' in
theoretical predictions of the expected Milky Way satellite
population.  Because our analysis is both based on and intended to be
compared with such theoretical predictions, we do not include these
systems in our quantitative analysis.

We thus conclude that the asymmetric distribution of Milky Way dwarfs
with latitude implies a realistic incompleteness in the Milky Way
dwarf census of $\sim 33\%$, with a range from 0$\%$ to 50$\%$
including variance due to Poisson noise, the fact that satellites may
not be randomly distributed, and the fact that the evidence discussed
above shows that there may be intrinsic asymmetry in the distribution.

\subsection{Incompleteness in the Outer Galactic Halo}
In this section, we use the known satellite population of M31 to
crudely estimate the number of Milky Way satellites similar to those
in the known population that may have been missed in past surveys.

Surveys for Local Group dwarf galaxies based on diffuse light have
been limited to central surface brightnesses brighter than 24 -- 25
mag arcsec$^{-2}$.  However, surveys for overdensities of resolved
stars have been able to identify five nearby ($\lesssim$ 110 kpc)
Milky Way dwarf satellites with $\mu_0$ fainter than 25 mag
arcsec$^{-2}$ (\citealp{wilson55,cannon77,irwin90,ibata94}).  Three of
these five were found by visual inspection, one was found
serendipitously, and one was found as an excess in total number
density of stars in scans of UKST plates.  Such surveys are less
sensitive to outer halo satellites (100 - 250 kpc)
because far fewer of their stars are resolved than in more nearby
satellites.  These surveys thus would have been unable to detect
distant ($>$ 100 kpc) dwarfs as faint as those detected more nearby.  Such
faint, outer halo systems thus lie in a ``blind spot'' of past
surveys.  \citet{kleyna97} did perform a systematic and automated
survey for resolved Milky Way companions over $25\%$ of the sky that
was sensitive to any of the known dwarfs to distances of 140 kpc.
However, 85$\%$ of the volume of the Milky Way's halo lies beyond 140
kpc, leaving open the possibility of undercounting in the outer halo.

\begin{figure}
\centerline{\psfig{file=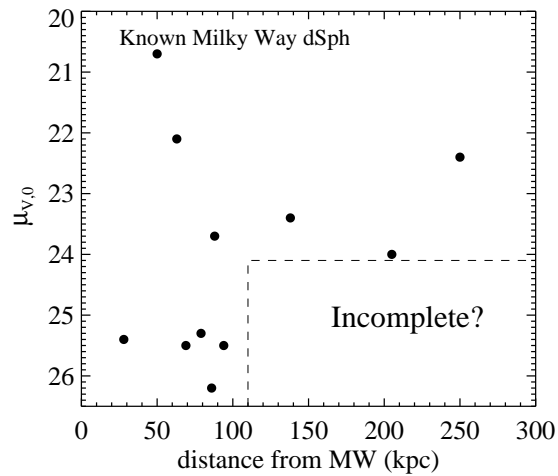,width=\hsize}}
\caption{The V-band central surface brightnesses of the Milky Way
dwarf companions as a function of their distance.  The dwarf census in
the boxed region may be incomplete because past surveys for Milky Way
dwarf satellites were less sensitive to galaxies in the outer
halo. Data from \citet{mateo98} and \citet{grebel03}. This figure is based on Figure 6 from
\citet{vdbergh99}. }
\label{fig:lgmudist}
\end{figure}

This observational bias may explain the notable dearth of Milky Way
dwarf galaxies more distant than 110 kpc with surface brightnesses
fainter than 24 mag arcsec$^{-2}$, as pointed out by \cite{vdbergh99}.
Figure~\ref{fig:lgmudist}, based on Figure 6 in \cite{vdbergh99},
shows the distribution of central surface brightnesses and
Galactocentric distances of known Milky Way
satellites. \citet{bellazzini96} also noted the apparent trend of
surface brightness with Galactocentric distance for Milky Way
satellites, not including the Magellanic Clouds.  They attributed the
trend to a true physical effect, due to the Galactic tidal field,
rather than to an observational bias.

\begin{figure}
\centerline{\psfig{file=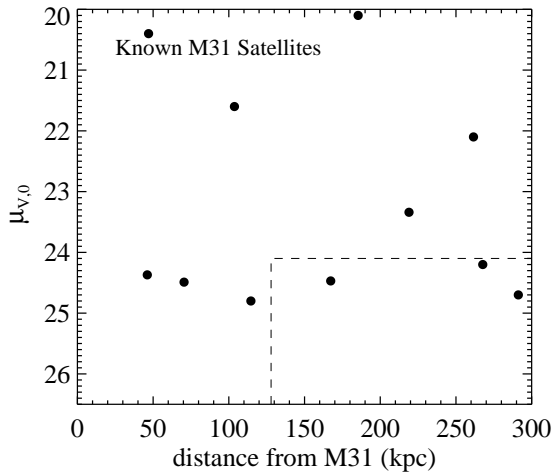,width=\hsize}}
\caption{The V-band central surface brightnesses of the M31 dwarf
companions as a function of their distance.  M32 is not on this plot
because it is brighter than the plotted range.  Data from
\citet{mateo98} and \citet{grebel03}.  The region corresponding to
that of a possible incompleteness in the known Milky Way population is
outlined by the dotted box.  There is no evidence for incompleteness
at low surface brightness and large radii, as was seen in the Milky
Way dwarf satellite distribution.  The cutoff seen around $\mu_{V,0}$
= 25 mag arcsec$^{-2}$ may be due to the limiting surface brightness
of surveys for Local Group dwarfs.}
\label{fig:m31mudist}
\end{figure}

 Using N-body simulations, \citet{mayer01} showed that tidal
stirring from the Galactic tidal field can serve to decrease the
surface brightness of dwarf galaxies.  Qualitatively, this effect
could result in a lack of ultra-low surface brightness satellites at
Galactocentric distances smaller than 100 kpc. It is possible to test
this alternative hypothesis by looking for a trend between surface
brightness and distance in M31 dwarf satellites.  Because M31 and its
satellites lie at a common distance and are detected by diffuse light,
one expects the satellites to be uniformly sampled with radial
distance from M31 and thus not to see a positive trend in their
$\mu_{V,0}$ with radial distance due to the bias described above.  One
instead expects to see a distance independent cutoff at the surface
brightness corresponding to the limiting sensitivity of existing sky
survey data.  Figure~\ref{fig:m31mudist} shows the distribution of
central surface brightness and Galactocentric distances of known M31
companions.  The lack of any radial trend in the dwarf companions to
M31 suggests that tidal interactions alone do not account for the
relative overabundance of dwarfs fainter than 24 mag arcsec$^{-2}$ in
the inner halo of the Milky Way.  However, this comparison is
inconclusive due to the fact that the 5 ultra-faint Milky Way
companions closer than 100 kpc may have been thus far undetectable
around M31. 


\begin{figure}
\centerline{\psfig{file=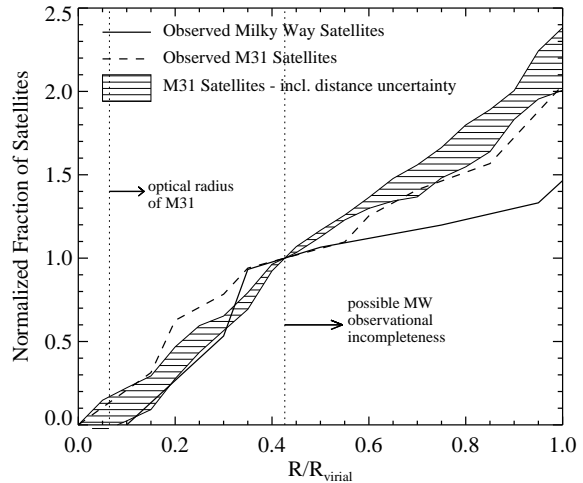,width=\hsize}}
\caption[Radial distributions of dwarf satellites to the Milky Way and
M31.]{Radial distributions of dwarf satellites to the Milky Way and
M31.  The radial distribution of M31 satellites, taking distance
measurement errors into account, is overplotted.  For reference, the
optical radius of M31, and the radius beyond which there is evidence
for observational incompleteness in the Milky Way dwarfs are also
overplotted.}
\label{fig:m31nvsr}
\end{figure}

Another way to evaluate the possibility of undercounting in the outer
halo is to compare the radial distribution of Milky Way satellites
with that of M31 satellites and see if they differ at large radii.
Figure~\ref{fig:m31nvsr} shows the radial distributions of both M31
and Milky Way satellites after normalizing the Galactocentric
distances of the dwarfs from each galaxy by their parent galaxy's virial
radius, R$_{vir}$ (258 kpc for the Milky Way and 280 kpc for M31 from
\citealt{klypin02}).  We also normalized the radial distributions to
the cumulative number of dwarfs within 0.43 R$_{virial}$, the most
uniformly sampled volume around the Milky Way. We overplot the optical
radius of M31, to show that obscuration by the disk of M31 does not
cause a substantial undersampling of its nearby dwarf satellites.

Figure~\ref{fig:m31nvsr} shows that M31 satellites are less biased to
lie at small radii than Milky Way satellites, as expected if Milky Way
satellites are undercounted at large radii. However, M31 satellites
have distance measurement uncertainties that range from 25 to 70 kpc
\citep{grebel03}, which may affect the utility of this comparison.  We
thus simulate the effect of M31 and M31 satellite distance
uncertainties on the measured radial distribution of M31 satellites.
To do this, we calculate a radial distribution for each of 1000
samples of M31 satellites with distances drawn from Gaussians with the
published distance uncertainties of each satellite. The distance to
M31 is also permitted to vary for each sample, according to its
distance measurement uncertainty \citep{stanek98}.  The 1-sigma range
of resulting radial distributions is overplotted on
Figure~\ref{fig:m31nvsr}.  This 'simulated' M31 radial distribution is
systematically less biased to small radii than the original M31
distribution, making it even less consistent with the observed Milky
Way distribution.  This unusual result stems from the fact that 7 of
M31's 12 satellites have measured distances within 30 kpc of M31's
distance.  Along a line of sight near M31, a satellite with
dist$_{sat-MW}$ $\sim$ dist$_{M31-MW}$ has the minimum
dist$_{M31-sat}$ possible.  Therefore, distance errors serve only to
increase dist$_{M31-sat}$ for a majority of M31 satellites.

A Kolmogorov--Smirnov test shows that Milky Way satellites are
formally consistent with being drawn from the same distribution as the
M31 dwarfs (not accounting for distance uncertainties).  This
similarity of the radial distributions suggests that the extent of
outer halo undercounting may not be substantial. However, the facts
that: {\it i)} the radial distribution of Milky Way satellites flattens
dramatically at the intermediate distances beyond which observational
bias would lead to an undercounting of Milky Way dwarfs and {\it ii)}
the two populations have such different surface brightness
distributions at the faint end, makes the case for inconsistency
stronger. Furthermore, distance uncertainties reduce the compatibility
of the Milky Way's and M31's radial distributions.

To quantitatively assess the possible number of missed Milky Way
satellites, we crudely estimate the number of additional dwarfs
necessary at $d \gtrsim$ 100 kpc for M31 and the MW to have the same
fraction of satellites within 0.43$R_{vir}$, the most uniformly
sampled volume around the Milky Way. If we assume that the Milky Way
population is uniformly sampled within 110 kpc and, like M31, that
half of Milky Way dwarfs lie beyond 110 kpc (0.43$R_{vir}$), we expect a total
of 15 -- 18 $\pm$ 4 Milky Way dwarfs.  The range in numbers is from
the range in radial distributions consistent with distance
uncertainties.  These upper and lower limits correspond to a range of
0 -- 11 undetected dwarfs more distant than 110 kpc.  However, these
numbers do not account for the possible incompleteness in the known
Milky Way population at low latitude, discussed in \S3.1.  Accounting
for a 33$\%$ incompleteness at low $b$ increases the above expected
total number of MW dwarfs to 25 -- 29 $\pm$ 5. Thus, we calculate a
total combined average incompleteness from both Galactic obscuration
and undersampling in the outer halo of $\sim 50\%$, with a possible
range of 0$\%$ to 66$\%$ incompleteness, including Poisson variation
and uncertainty in the true distribution of satellites with latitude.
We emphasize that these numbers only account for dwarfs with
properties similar to the known population and do not extrapolate to a
population of even fainter dwarfs, should they exist.

The robustness of these estimated numbers is affected by the small
numbers of dwarfs (and hence the large possible fluctuations from the
underlying populations) and the fact that the known population of M31
may not represent the underlying distribution of dwarfs down to 26 mag
arcsec$^{-2}$.  However, these numbers are simply intended to
underscore the necessity of considering incompleteness when matching
models to observations, and to provide a prediction that may be
testable by the \citet{willman02} survey for Milky Way dwarf
companions.  Several marginal cases of additional Milky Way dwarf
companions have been identified within 110 kpc, such as the Monoceros
stream (Yanny et al. 2003, among others) and $\omega$ Cen
\citep{lee99,dinescu99,majewski00}.  Including these sources in the
analysis would exacerbate the discrepancy in the radial distributions
and result in a larger predicted possible number of undetected
satellites.

The quantitative predictions for the number of Milky Way dwarf
satellites with properties similar to the known population are
summarized in Table~\ref{tab:predictions}.  The first column gives the
distribution the Milky Way was compared to (in this case, M31).
The next 4 columns give the predicted values for: the total number of
MW dwarfs assuming no incompleteness at low latitude, the total number
of MW dwarfs assuming a uniform distribution in latitude, the
undetected number of dwarfs beyond 110 kpc assuming no incompleteness
at low latitude, and the total undetected number of dwarfs (both at
low b and in the outer halo) assuming a uniform distribution.

\begin{table}
\renewcommand{\arraystretch}{1.2}
\begin{tabular}{ccccc}
\hline
method & tot & tot$_{b,corr}$ & undet & undet$_{b,corr}$  \\ \hline
M31 & 15 - 18 $\pm$ 4 & 25 - 29 $\pm$ 5 & 4 - 7 $\pm$ 4 & 14 - 18 $\pm$ 5 \\
oldest & 14 $\pm$ 4 & 22 $\pm$ 5 & 3 $\pm$ 4 & 11 $\pm$ 5   \\
highest v$_c$ &  20 $\pm$ 4  &  33 $\pm$ 6  &  9 $\pm$ 4  &  22 $\pm$ 4  \\ \hline
\end{tabular}
\caption[]{Predicted Number of Milky Way Dwarfs With Properties Similar to the Known Population}
\label{tab:predictions}
\end{table}

\section{The Spatial Distribution of Galaxy Sub-Halos in $\Lambda$CDM Simulations}
In this section, we compare the observed radial distributions of Milky
Way and M31 satellite galaxies with the radial distributions predicted
by two different simplistic substructure models applied to the Milky
Way-like galaxy from a the high resolution $\Lambda$CDM cosmological
simulation described in \S2.  We use these two models to highlight the
impact of a possible incompleteness on the robustness of such
comparisons.  We then estimate the number of additional Milky Way
dwarfs possible at $d \gtrsim$ 110 kpc for the observed and the model
distributions to be consistent.

Following Taylor et al (2003), we use either the oldest or the highest
v$_c$ sub-halos to characterize two popular substructure scenarios.
The oldest sub-halos would preferentially be observable as luminous
satellite galaxies in a scenario where reionization is the dominant
physics that effects the observability of low mass sub-halos.  In this
scenario, low mass sub-halos that form after reionization cannot
accrete as much neutral gas as halos that formed before reionization,
if they can accrete any at all, making it difficult for sub-halos that
collapse after reionization to ever form stars.  On the other hand,
\citet{stoehr02} and \citet{hayashi03} recently suggested that the
highest v$_c$ sub-halos (at z = 0) may preferentially be observable as
Milky Way satellites, if the circular velocities of the observed Milky
Way satellites have been grossly underestimated.  In that scenario, a
combination of reionization, feedback, and tidal effects could have
rendered all of the less massive sub-halos thus far
unobservable. Although these two models clearly are not complete
descriptions of the physics that affects sub-halo luminosity, they are
sufficient for the purposes described above.

\subsection{Identifying Oldest and Highest v$_c$ Sub-halos}

To select the oldest sub-halos, we reconstruct the trajectory of each
sub-halo within R$_{vir}$ of the galaxy at z = 0 back to z = 9.7.  We
used the mass history of each sub-halo to interpolate the time at
which each contained both 50$\%$ and 25$\%$ of its peak mass. A
sub-halo was the progenitor, P$_b$, a of a sub-halo, S$_a$ if it
contained the highest fraction of the number of particles in S$_a$.
If multiple halos contained $>$ 10$\%$ of S$_a$'s particles, we
selected P$_b$ as the sub-halo that contributed the highest fraction
of its 10 most bound particles to S$_a$, following \citet{delucia04}.

We defined the highest v$_c$ sub-halos as those with the highest peak
circular velocities, $v_{peak}$, at z = 0.  The circular velocities
are simply determined by $v_c = (GM/r)^{0.5}$, out to each sub-halo's
tidal radius.  We find that the 15 sub-halos with the highest
$v_{peak}$s at z = 0 include the 12 sub-halos with the highest masses
along their past trajectory.

\subsection{Radial Distribution of the Oldest and Highest v$_c$ Sub-Halos}
\begin{figure}
\centerline{\psfig{file=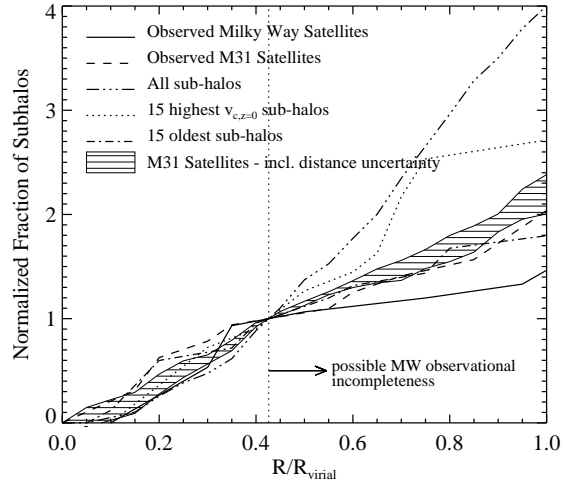,width=\hsize}}
\caption{The radial distributions of: dark matter sub-halos of a 3
$\times 10^{12} M_{\odot}$ galaxy in a $\Lambda$CDM dark matter only
cosmological simulation \citep{governato03}, the 15 highest v$_c$
sub-halos, the 15 oldest sub-halos, and the known Milky Way and M31
dwarf satellites.  We also overplot the spread, due to distance
uncertainties, of M31 radial distributions (see \S3.2).  The highest
v$_c$ sub-halos are defined as those with the highest v$_{peak}$ at z
= 0, and the oldest sub-halos are defined as those that accreted
50$\%$ of their peak mass at the earliest time.}
\label{fig:numvsdistlog}
\end{figure}

Figure~\ref{fig:numvsdistlog} shows the radial distributions of the
entire DMgal sub-halo population, the oldest and highest v$_{c,z=0}$
sub-halos, and the dwarf populations of the Milky Way and M31.  The
spread in M31 radial distributions due to distance uncertainties is
also overplotted (see \S3.2).  We used the sub-halos that accreted
50$\%$ of their peak mass at the earliest time to define the oldest
population.  Again, we have normalized the distances by the virial
radius of the parent halo to account for differences in the size and
mass of the Milky Way, M31 and DMgal.  We have also normalized the
radial distributions to the cumulative number of dwarfs within 0.43
R$_{virial}$, the most uniformly sampled volume around the Milky Way.
Due to small numbers, KS tests show that all of the plotted
distributions, except for that of the entire sub-halo distribution,
are at least marginally consistent with each other.  This consistency
highlights a potential difficulty in using the radial distribution of
a single population to rule out models.  However, some of the
distributions are much more similar than others, which we discuss
below.

This figure shows that sub-halos with the highest peak velocities
at z = 0 are biased to lie at smaller radii than the overall sub-halo
population.  A KS test of the two distributions shows they are not
inconsistent with being drawn from the overall sub-halo population with
$\sim$30$\%$ certainty. This radial bias was also found by
\citet{taylor03} and \citet{kravtsov04} for galaxy sub-halos, and
\citet{governato01} and \citet{diemand04} for the highest v$_{peak}$
sub-halos of galaxy clusters.  To understand this radial bias,
consider that the radial distribution of the highest v$_{peak}$
sub-halos is very similar to that of the sub-halos with the
highest mass along their past trajectories, as stated in \S4.1.  The
sub-halos surviving at small Galactocentric distances at z = 0 are
those that were, on average, the most robust to tidal disruption.
Sub-halos with the highest masses in the past were both more robust to
tidal disruption, and would have experienced dynamical friction that
would have reduced their apocenter distances.

The fact that the highest v$_c$ sub-halos have a distribution that is
much less biased to small radii than that of even M31's satellites,
suggests that the Stoehr/Hayashi model may not be correct.  A recent
dynamical study by \citet{stelios04} reaches the same conclusion.
Nevertheless, to evaluate the possibility of using radial
distributions to distinguish between substructure models, we compute
the number of undetected outer halo Milky Way satellites necessary for
the same fraction of them and of the highest v$_c$ sub-halos to lie
within 0.43$R_{vir}$ (as we did in \S3.2).  The resulting total number
of dwarfs similar to the known population, in the highest v$_c$
sub-halo model, ranges from 20 -- 33, depending on the assumed
latitude distribution of the satellites.  These numbers are summarized
in Table 2.

Similar to the highest v$_c$ sub-halos, the oldest sub-halos are
biased to lie at smaller radii than the overall sub-halo population,
but even more so.  In fact, there is a striking similarity between the
radial distribution of the oldest sub-halos and the observed satellite
galaxies, particularly of M31's. The scatter in the M31 satellites'
radial distribution, due to distance uncertainties, reduces its
similarity to that of the oldest sub-halos.  However, a KS test shows
that, at worst, they are consistent with each other at $> 80\%$, which
is more than any of the other distributions.  The radial distribution
of the oldest sub-halos defined by the time they had accreted 25$\%$
of their peak mass, rather than 50$\%$, also matches the plotted
distribution very closely.  We again compute the number of undetected
Milky Way satellites necessary in the outer halo for the fraction of
Milky Way satellites within 0.43$R_{vir}$ to exactly match that of the
oldest sub-halos. The numbers are summarized in Table 2.

The close match between the radial distribution of the oldest
sub-halos and that of M31 seems to indicate that reionization is a
primary factor effecting the observability of sub-halos.  However,
\citet{kravtsov04} used a more detailed approach and found that the
observable properties of Galactic satellites are primarily a function
of the physics of galaxy formation, rather than reionization.  This
different result demonstrates that cosmic scatter intrinsic to both
observed and simulated satellite distributions may be a large enough
effect to make it difficult to distinguish between substructure models
solely using a small sample of radial distributions.  This potential
pitfall is reflected in the fact that the numbers for the 3 different
models in Table 1 are all consistent within their Poisson errors.

\section{Predictions for a New Dwarf Galaxy Survey}
In this section, we determine if any of the above predictions for the
number of undetected outer halo Milky Way satellites similar to those
known will be testable by the new SDSS search for resolved dwarf
galaxy companions to the Milky Way \citep{willman02}.  To do this, we
calculate the number of undetected dwarfs similar to the known
population, but more distant than 110 kpc, that may lie in the SDSS
area under various sets of assumptions.  Because SDSS only observes at
$b > 30$, it cannot constrain incompleteness at low $b$.  The number
of predicted dwarfs in the surveyed area is independent of whether we
assume a uniform distribution in latitude or assume that the observed
distribution in latitude accurately reflects the underlying
distribution.  We thus compute the number of undetected dwarfs as:

\begin{equation}
f_{|b|>30} \cdot f_{obs,|b|>30} \cdot n_{undet,nocorr},
\end{equation}

\noindent where $f_{|b|>30}$ is the fraction of satellites observed to
lie above $|b|$ = 30, $f_{obs,|b|>30}$ is the fraction of $|b| > 30$
sky that SDSS will image, and $n_{undet,nocorr}$ is the number of
predicted undetected galaxies with no latitude correction.

When complete, the SDSS will cover $\sim$ 25$\%$ of the entire
sky. Based on the numbers in Table~\ref{tab:predictions}, we expect a
total of only 2-3 $\pm$ 1 undetected outer halo dwarfs from the M31
`model', a total of 1 $\pm$ 1 dwarf from the oldest sub-halo model,
and a total of 4 $\pm$ 2 dwarfs in the highest v$_c$ sub-halo model.
In the event of a null detection, the SDSS coverage will not be
sufficient to definitively assess whether the underlying radial
distribution of MW dwarfs is exactly consistent with any of these
three models.  However, the detection of a substantial number of outer
halo dwarfs would call the ``oldest sub-halo'' (reionization) model
into question.  In a future paper, we will assess the number of dwarfs
fainter than those in the known population, as predicted by various
substructure models, that the new SDSS search should be sensitive to.

\section{Conclusion}
In this paper, we have reviewed evidence for incompleteness in the
known Milky Way dwarf satellites with properties similar to those in
the known population.  This possible incompleteness is due to past
observational bias against detecting Milky Way satellites at low
latitude and in the outer Galactic halo.  Although the level of
incompleteness is very uncertain, the fact that an empirical
extrapolation from the M31 distribution yields an average total number
of MW dwarfs that is 1.5 -- 2 times the known population shows that
incompleteness needs to be taken seriously when comparing to models of
dwarf galaxy formation.

We used the oldest and highest v$_c$ sub-halos of a simulated Milky
Way-like galaxy to demonstrate how radial distributions may be used to
distinguish between proposed models of dwarf galaxy
formation. However, KS tests comparing the radial distributions of the
Milky Way, M31, and the oldest sub-halos and the highest v$_c$
sub-halos in simulations show that they are all at least marginally
consistent with each other. Interestingly, the M31 distribution is
consistent with the oldest sub-halo distribution at $> 95\%$,
suggesting that reionization may have a substantial effect on the
observabilitiy of sub-halos (however, see Kravtsov et al. 2004).
However, small numbers and cosmic scatter permit at least a marginal
consistency between a wide range of observations and models. It is
thus difficult at present to use radial distributions alone to clearly
distinguish between substructure models, although they certainly
provide a complimentary test of model predictions. Though faint galaxy
membership in other groups is currently controversial, when the
satellite populations of galaxies in nearby groups are known more
precisely, their radial distributions will provide a stronger
discriminant between models. Likewise, a large ensemble of high
resolution simulations will allow a more robust assessment of the
effects of cosmic scatter and small numbers on the predicted satellite
population.

The crude arguments presented in this paper result in predicted total
numbers of dwarfs that range from 1 -- 3 times the known number, with
the most realistic estimates producing an incompleteness in the Milky
Way dwarf census of up to 50$\%$. The exact level of incompleteness is
strongly dependent on the distribution of Milky Way satellites with
latitude.  If the Milky Way census is incomplete at the level of
50$\%$ or more, then many existing models underpredict the number of
luminous Milky Way dwarfs.  In particular, models with suppressed
small scale power would not produce enough luminous dwarf galaxies, as
has already been suggested by \citet{chiu01}.  Note that our derived
``total incompleteness'' only accounts for dwarfs with properties
similar to those known, not any fainter dwarfs, should they exist.

Currently, the largest uncertainty in the known Milky Way population
is the underlying distribution of dwarfs with latitude.  Before the
Milky Way satellite population can yield a meaningful comparison with
substructure models, this uncertainty needs to be investigated with
more detail than presented in this paper. The average incompleteness
in the current census, assuming a uniform distribution in latitude and
no other incompleteness, is 33$\%$.  Any survey sensitive to faint
dwarf satellites at low latitude would thus place very valuable
constraints on the Local Galaxy luminosity function.


\section*{Acknowledgments}
We would like to acknowledge Andrey Kravtsov for conversations that
substantially contributed to this paper. We would also like to thank
Mike Blanton and Andreas Berlind for helpful conversations.  BW was
supported in part by NSF grant AST-0098557, NSF grant AST-0205413, and
the UW Royalty Research Fund.  JJD and BW were partially supported
through NSF grant CAREER AST-0239683 and through the Alfred P.\ Sloan
Foundation. FG is a Brooks fellow and was supported in part by NSF
grant AST-0098557 at the University of Washington. DR acknowledges
support from the NASA Graduate Student Researchers Program and from
PPARC.  TRQ acknowledges support from NSF grant AST-0098557 and NSF
grant AST-0205413.




\end{document}